\documentclass{iaus}
\usepackage{graphicx}

\title[Dark matter in early-type galaxies with PNe] 
{Dark-Matter Content of Early-Type Galaxies with Planetary Nebulae}

\author[Napolitano et al.]   
{N.R. Napolitano$^1$,
A.J. Romanowsky$^{2}$, L. Coccato$^{3}$, M. Capaccioli$^{4,5}$ N.G. Douglas$^{6}$, E. Noordermeer$^{7}$, M.R. Merrifield$^{7}$, K. Kuijken$^{8}$, M. Arnaboldi$^{9}$, O. Gerhard$^{3}$, K.C. Freeman$^{10}$, F. De
Lorenzi$^{3}$, P. Das$^{3}$}

\affiliation{$^1$INAF-Observatory of Capodimonte, Napoli, Italy email:napolita@na.astro.it\\[\affilskip]
$^2$ Departamento de F\'{i}sica, Universidad de Concepci\'{o}n, Chile,
$^3$ MPE - Garching, Germany\\[\affilskip]
$^4$ Department of Physics, University ``Federico II'' Naples, 
$^5$ INAF - VSTceN, Naples, Italy\\[\affilskip]
$^6$ Kapteyn Institute, Groningen, the Netherlands, 
$^7$ University of Nottingham, UK\\[\affilskip]
$^8$ University of Leiden, the Netherlands, 
$^9$ ESO - Garching, Germany\\[\affilskip]
$^{10}$ RSAA, Mt. Stromlo Observatory, Australia\\[\affilskip]}

\pubyear{2007}
\volume{244}  
\pagerange{349--354}
\date{01-09-2007}
\jname{Dark Galaxies and Lost Baryons}
\editors{J. I. Davies \& M. J. Disney, eds.}
\begin{document}

\maketitle

\begin{abstract}
We examine the dark matter properties of nearby early-type galaxies using planetary nebulae (PNe) as mass probes. We have designed a specialised instrument, the Planetary Nebula Spectrograph (PN.S) operating at the William Herschel telescope, with the purpose of measuring PN velocities with best efficiency. The primary scientific objective of this custom-built instrument is the study of the PN kinematics in 12 ordinary round galaxies. Preliminary results showing a dearth of dark matter in ordinary galaxies (Romanowsky et al. 2003) are now confirmed by the first complete PN.S datasets. On the other hand early-type galaxies with a ``regular'' dark matter content are starting to be observed among the brighter PN.S target sample, thus confirming a correlation between the global dark-to-luminous mass virial ratio ($f_{\rm DM}=M_{\rm DM}/M_\star$) and the galaxy luminosity and mass.     
\end{abstract}

\firstsection 
\section{Introduction}
The currently accepted cosmological paradigm, the so-called concordance $\Lambda$CDM model, i.e. cold dark matter (CDM) plus a cosmological constant ($\Lambda$), has been overwhelmingly successful in describing the formation and evolution of structures in the Universe but there remain many observational discrepancies on galaxy and cluster scales that call for a critical verification.
While the flat rotation curves observed since the 1970s in spiral galaxies have been one of the prime factors in support of the existence of dark matter (Einasto et al.1974; Faber\&Gallagher 1979), studies of late-type galaxies suggest, in some cases (e.g. low surface brightness galaxies), soft cores and relative low concentration density profiles (Gentile et al. 2004; de Blok 2005
) at odd with $\Lambda$CDM inner density cusps (Navarro et al. 1997, NFW hereafter;  Moore et al. 1999).
Given the long-standing dark matter puzzles in disk galaxies, it is obviously important to also examine the mass content of the second major galaxy type, the early-types (ETGs: ellipticals and S0). The mass content of these systems, which are generally free of cold gas and have their stars moving in random directions, is more difficult to measure than in disk galaxies. The main difficulty is observational, since the only ubiquitous mass tracers in ETGs, i.e. stars, can be investigated through their integrated light (spectroscopy) which is hard to be measured with sufficient signal-to-noise outside $\sim 2 R_{\rm eff}$ (Kronawitter et al. 2001, Samurovic \& Danziger 2005). A suitable option is the X-ray emission from hot gas (Paolillo et al. 2003, O'Sullivan \& Ponman 2004, Humphrey et al. 2006), but this turns out to be difficult to be measured around faint galaxies (Fukazawa et al. 2006). Very promising mass probes are also globular clusters (Grillmair et al. 1994, Richtler et al. 2004, Puzia et al. 2004) and more importantly a particular class of stars, planetary nebulae (Ciardullo et al. 1993, Arnaboldi et al. 1998, Napolitano et al. 2001, Peng et al. 2004). 
None of these techniques is free of limitation, thus after years of investigations there are relatively few ellipticals with a dark matter halo even confirmed and much less shown to have the dark-matter properties expected with $\Lambda$CDM.\\ 
We will concentrate here on the observational tests on ETGs allowed by PNe which have so far ascertained the presence of both massive (Napolitano et al. 2002, Peng et al. 2004) and weak dark haloes (Mendez et al. 2001, Romanowsky et al. 2003) which can be either interpreted as a variation of the star formation efficiency or as the reflection of a concentration problem similar to the one found in late-types (Romanowsky et al. 2003, Napolitano et al. 2005). Yet it is still unclear if the discrepancies can be traced to observational/modeling problems, to a poor knowledge of the baryonic physics (in N-body simulations predicting the CDM distribution), or to a failure of the $\Lambda$CDM paradigm.

\vspace{-0.4cm}
\section{Measuring galaxy kinematics with the Planetary Nebula Spectrograph}\label{pns}
Planetary Nebulae (PNe) have been proven to be excellent tracers for the dynamics of outer regions of ETGs. Through their powerful [OIII] emission at 5007 \AA,  they are easily detectable and their radial velocity measurable also in halo regions of distant galaxies. From 2001 a dedicated instrument to the PN kinematics in galaxy systems, the {\it Planetary Nebula Spectrograph} (PN.S), is operating at the 4m Herschel Telescope on La Palma. The PN.S is specifically designed for the measurement of kinematics of
extragalactic planetary nebulae. It allows this type of observations to be
carried out a factor of ten more efficiently than possible so far, mainly thanks to its design which allows the PNe to be discovered and their spectra
measured in a single observation through the ``Counter-Dispersed Imaging"
technique (Douglas et al. 2002). The primary program of the PN.S is to survey a dozen bright ($m_B\leq$~12), round (E0--E2), nearby ($D\leq 25$~Mpc) ellipticals with a large range of stellar light parameters (luminosity, concentration, shape),
rotational importance, and environment, with the goal of observing 100--400 PN velocities in each. The final reduction pipeline of the instrument data, the velocity and photometric calibration procedures plus comparison of the PN.S velocities of the first complete datasample, the one of NGC~3379, with external independent PN catalogues has been discussed in Douglas et al. (2007): we address the reader to this paper for details on the data-reduction and standard kinematical analysis and some basics dynamics of NGC~3379.   
\begin{table}\def~{\hphantom{0}}
  \begin{center}
  \caption{Galaxy properties: the PN.S samples and main Jeans model parameters}
  \label{tab:kd}
  \begin{tabular}{lccccccc}\hline
      $ID$  & D (Mpc)  & $M_B$ & N. PNe& $c_{\rm vir}$ & $M_{\rm vir}$(10$^{11} M_\odot$) & $f_{\rm DM}$ & $\beta (4R_{\rm eff})$ \\\hline
       NGC 821 & 25.5 & -20.5 & 130& 5 & 6.0 & 6 & -0.1\\
       NGC 3379  & 9.8 & -20  & 191 &4 & 4.2 & 5 & 0.3\\
       NGC 4494 & 15.9 & -20.5& 255 & 4 & 9.1 & 9 & 0.3\\
       NGC 5846 & 26.3& -21.3 & 140 & 10 & 80 & 16 & 0.0 \\\hline
  \end{tabular}
 \end{center}
\end{table}

\vspace{-0.7cm}
\section{Falling versus flat dispersion profiles}\label{sec:trapmode}
Preliminary dataset obtained with the PN.S for three ``ordinary'' elliptical galaxies have been presented in Romanowsky et al. 2003 (R+03 hereafter). The velocity dispersion (VD, herafter) profile of these three systems showed a pseudo-Keplerian decline that was more consistent with constant mass-to-light ratio systems rather than dark matter dominated systems for which a flat VD profile was expected. 

This unexpected result has produced different interpretations either in the $\Lambda$CDM framework (Dekel et al. 2005 $=$ D+05) or in MOND theory (Milgrom \& Sanders 2003). In particular, D+05 address very radial stellar orbits and projection effects in order to explain declining VD profiles.  

Napolitano et al. (2005, N+05 hereafter) made predictions of gradients of mass-to-light ratios in ETGs and have ascertained that ``quasi-constant M/L'' are indeed expected within $\Lambda$CDM, although the R+03 sample shows $M/L$ gradients which are too low and conflicting with acceptable star formation efficiency and baryon fraction.
Low M/L gradients mirror a generalised trend of the global dark-to-luminous mass virial ratio ($f_{\rm DM}=M_{\rm DM}/M_\star$) with stellar mass/luminosity: brightest galaxies have a larger fraction of dark-
to-luminous matter, $f_{\rm DM}=$ with respect to fainter galaxies ($M_B>-20.5$), the trend being possibly reversed toward dwarf galaxies. 
Here we characterise the combined VD profiles of the R+03 galaxy sample, i.e. NGC~821, NGC~3379 and NGC~4494, for which, after 4 years, we have now the final PN datasets (see Table 1 for the final number of PNe observed) and present very preliminary dynamical estimate of their dark matter content as measured through their $f_{\rm DM}$. 

In Fig. 1 (left panel), with open symbols, we show the combined VD profile of the three galaxies after having scaled (with respect to $R_{\rm eff}$) and normalised (with respect to the central velocity dispersion) the individual galaxy curves. These have been obtained as azimuthally averaged $V_{\rm rms}=\sqrt{v^2+\sigma^2}$ within radial bins following the galaxy isophotes distribution for all galaxies. It is easy to recognise the pseudo-Keplerian fall consistent with the one observed in the preliminary datasets as in R+03. In the same figure we also show a collection of long-slit data from the same galaxies as empty stars which demonstrate that the PN VD profiles lie on the extention of the stellar kinematics of the central regions, thus allowing to continuously map the galaxy kinematics out to many effective radii. 
Beside their declining stellar VDs, these galaxies share other common properties like intermediate luminosities ($\sim L_*$, see Table 1) and the fact of being fast rotators (Cappellari et al. 2006), disky/cuspy systems (N+05). 

In the same panel of Fig. 1, we also show for comparison two galaxies, NGC~4374 and NGC~5846 (full symbols), for which we have the PN.S observational program lately completed and final PN catalogues done (see N. PNe in  Table 1). Here the velocity dispersion profiles have been normalised to the average central VD value of the faint galaxies in order to mark the kinematical differences between the two samples; long-slit kinematics are also shown as full stars, accordingly. 
Differently from the R+03 sample, these galaxies show markedly flat VD profiles which well contrasts with the pseudo-Keplerian fall of the former ones, suggesting the presence of a significant dark halo. These two ``dark-matter dominated'' galaxies mark differences with respect to the R+03 sample for many other reasons: they are brighter ($M_B<-20.9$), slow-rotators (Cappellari et al. 2006) and boxy/cored galaxies (N+05). This dichotomy on many structural galaxy properties is suggestive of some connection with galaxy formation history as discussed elsewhere (Capaccioli et al. 2002, N+05).
\begin{figure}
\centering
\resizebox{\hsize}{!}{\rotatebox[]{0} {\includegraphics[height=5.cm,width=5.cm]{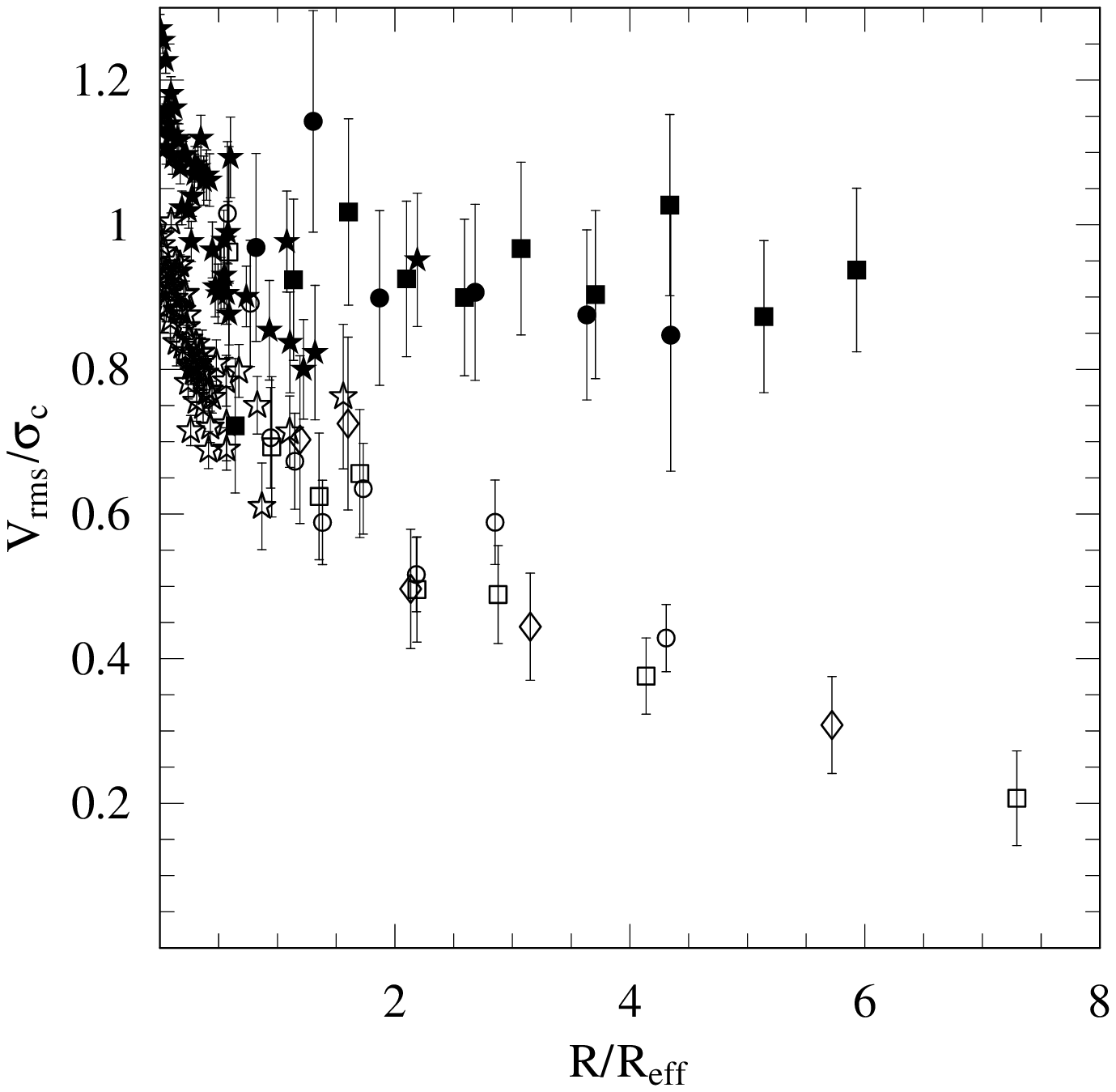}}\rotatebox[]{-90} { \includegraphics[height=5.1cm,width=5.1cm]{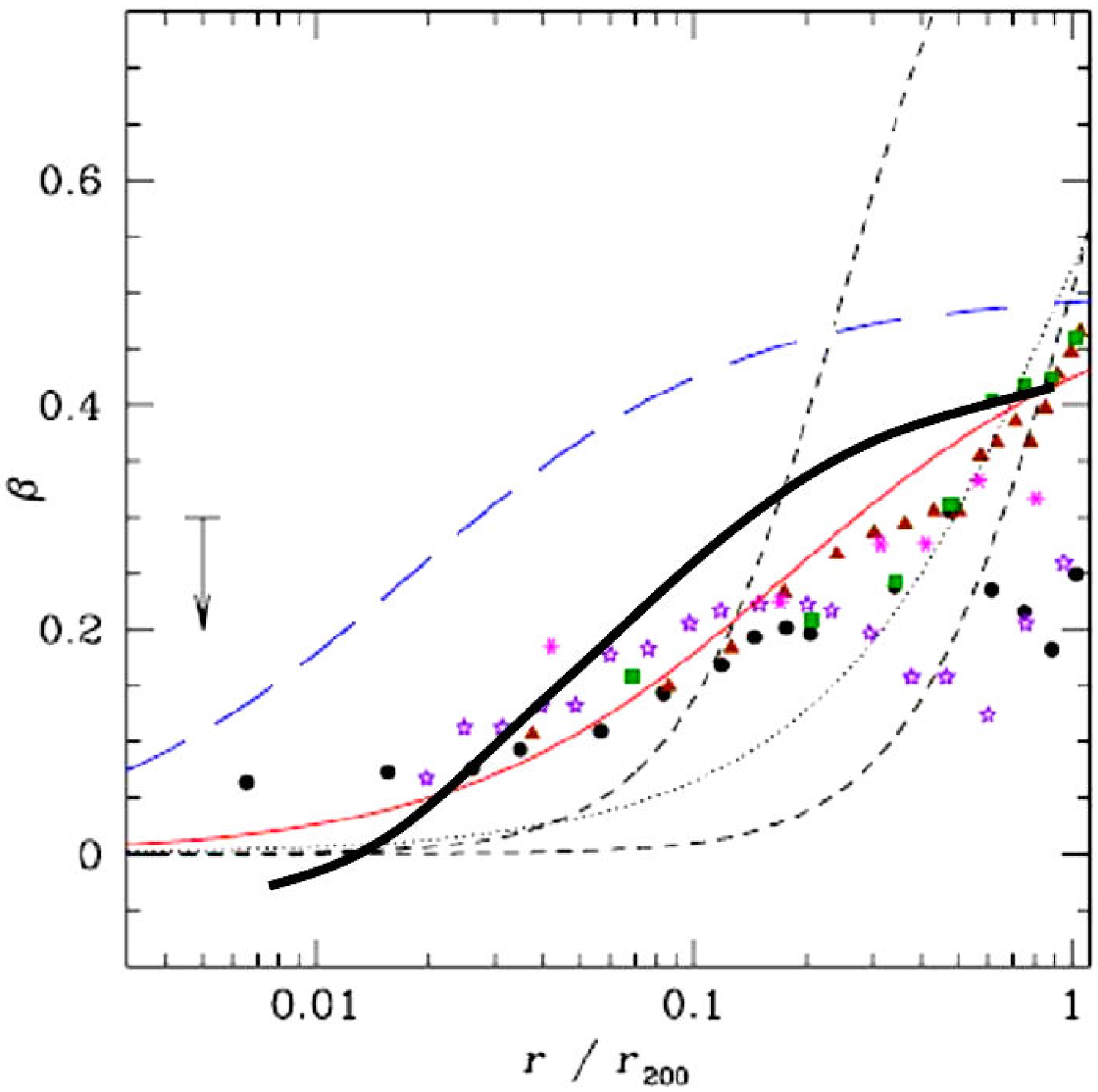}}}
  \caption{Left: Combined velocity dispersion profile of 5 galaxies from the PN.S primary target list (see text). The pseudo-Keplerian fall of the faint sample (NGC~821, NGC~3379, NGC~4494, open symbols) is evident against flat profiles of the brighter systems (NGC~4374 and NGC~5486, full symbols). Long-slit data from the inner integrated light are shown as star symbols (open and full accordingly). Right: the anisotropy profile corresponding to our best fit to the PN dispersion profile for NGC~4494 (black solid line) is overplotted to Fig. 2 from Mamon and Lokas 2005 where they show with symbols results from different N-body simulations and with curves analytical galaxy anisotropy profiles (long dashed matches the D+05 merger remnants).}
\end{figure}

Basic Jeans analysis was performed to substantiate the dark matter content. We assumed spherical symmetry (well motivated by the round appearence of the galaxy sample discussed here) and we have fit the $V_{\rm rms}$ by solving the radial Jeans equation and projecting the solution along the line-of-sight. The dark halo is modeled as a standard NFW using $c_{\rm vir}$ and $M_{\rm vir}$ (the dark halo concentration and virial mass respectively defined for $\rho/\rho_{\rm crit}=100$) as free parameters. We assumed either a constant or radial dependence for anisotropy parameter, $\beta=1-\sigma_{\theta}^2/\sigma_{r}^2$, in order to investigate the effect of tangential ($\beta<0$) and radial ($\beta>0$) orbits on the mass estimates. The main results for four galaxies (NGC~4374 is not reported here) are shown in Table 1. 

As anticipated, NGC~5846 is the only galaxy showing a ``regular'' dark matter halo, having the concentration and virial mass consistent with the $c_{\rm vir}-M_{\rm vir}$ relation expected from cosmological simulations (Bullock et al. 2001, N+05). The faint ``ordinary'' sample instead shows concentrations which are systematically lower than predicted from simulations. Another remarkable result is that the virial mass-to-luminous matter ratio, $f_{\rm DM}$ is clearly an increasing function of the galaxy stellar mass, which confirms a smooth trend shown in N+05 and discussed in Napolitano et al. (2006). This trend can be interpreted in terms of global star formation efficiency, $\epsilon_{\rm SF}=M_{\star}/M_{\rm bar}$, i.e. the fraction of baryonic mass $M_{\rm bar}$ cooled in stars, assuming baryon conservation, such that  $\epsilon_{\rm SF}=4.9/f_{\mathrm d}$  (see N+05 for further details). More massive systems are globally less efficient in converting gas to stars than galaxies around $L_*$, like the ones in the ``ordinary'' PN.S sample, which represent the most efficient ones placing themselves in the maximum (minimum) of the $\epsilon_{\rm SF}$ ($f_{\rm DM}$) run with $M_{\star}$. Indeed the trend is expected to be reversed moving toward the dwarf galaxies.
The fact that in our analysis we are considering appropriate orbital anisotropy, consistent with predictions from cosmological simulations (see Fig. 1 right panel and Table 1), allows as to exclude that the trend of the dark matter fraction within the analysed galaxies can be the reflection of an anisotropy sequence with the mass.   

%

%

\vspace{-0.4cm}
\section{Conclusions}\label{sec:concl}
We have shown that there is a strong indication of a dichotomy between the kinematic behaviour of the first PN complete datasets obtained with the PN.S. Faint ``ordinary'', fast-rotating, discy/cuspy, early-type systems seem to show declining velocity dispersion profiles out to very large distance from the galaxy centres, while bright, slow-rotating, boxy/cored systems have flat velocity dispersion profiles.  
This corresponds to a different total dark-to-luminous matter virial ratio, which is larger for brighter systems with respect to fainter ones. The only galaxy for which we have a preliminary Jeans analysis in the PN.S target sample, NGC~5846, shows a DM distribution which fairly follows a ``regular'' NFW density profile since the $c_{\rm vir}$ and $M_{\rm vir}$ are consistent with the expectation of N-body simulations (Bullock et al. 2001, N+05). The faint sample, on the contrary, has $c_{\rm vir}$ which are a factor of two lower than predicted from $\Lambda CDM$.

All these results seem to support the scenario discussed in Napolitano et al. (2005) where we interpreted the dichotomy in the global structure galaxy parameters as a correlation between dark matter content in galaxies and the mass. This is the consequence of a broader trend of the global efficiency of the star formation with the mass (Benson et al. 2000, Dekel \& Birnboim 2006) which suggests a connection between the observed dark matter properties of early-type galaxies and their formation history.  
\vspace{-0.2cm}
\begin{acknowledgements}
NRN has been granted by CORDIS with an FP6-Reintegration Grant, MERG-FP6-CT-2005-014774, co-funded by INAF.
\end{acknowledgements}

\end{document}